\documentclass[prl,twocolumn,superscriptaddress,showpacs, letterpaper]{revtex4-1}

\usepackage{graphicx} 
\usepackage{dcolumn}   % needed for some tables
\usepackage{float}
\usepackage{wrapfig}
\usepackage{verbatim}
\usepackage{amsmath}
\usepackage{bm}        % for math]]
\usepackage{amssymb}
\usepackage{amsfonts}
\usepackage{accents}
\usepackage{fancyhdr}
\usepackage{euscript}
\usepackage{blkarray}
\usepackage{multirow}
\usepackage{tikz}
\usepackage[nolist,nohyperlinks]{acronym}
\usepackage{enumitem}
\newcommand{\bea}{\begin{eqnarray}}
\newcommand{\eea}{\end{eqnarray}}
\newcommand{\be}{\begin{equation}}

\newcommand{\ee}{\end{equation}}
\newcommand{\ba}{\begin{align}}
\newcommand{\ea}{\end{align}}
\newcommand{\bi}{\begin{itemize}}
\newcommand{\ei}{\end{itemize}}
\newcommand{\bt}{\begin{tabular}}
\newcommand{\e}{\end{tabular}}

\newcommand{\vl}{\kern-\arraycolsep & \kern-\arraycolsep}
\newcommand{\VL}{\kern-\arraycolsep\strut\vrule &\kern-\arraycolsep}

\begin{document}
\title{Layered Topological Crystalline Insulators}
\author{Youngkuk Kim}
\affiliation{The Makineni Theoretical Laboratories, Department of Chemistry, University of Pennsylvania, Philadelphia, Pennsylvania 19104-6323, USA}
\author{C. L. Kane}
\affiliation{Department of Physics and Astronomy, University of Pennsylvania, Philadelphia, Pennsylvania 19104-6396, USA}
\author{E. J. Mele}
\affiliation{Department of Physics and Astronomy, University of Pennsylvania, Philadelphia, Pennsylvania 19104-6396, USA}
\author{Andrew M. Rappe}
\affiliation{The Makineni Theoretical Laboratories, Department of Chemistry, University of Pennsylvania, Philadelphia, Pennsylvania 19104-6323, USA}
\date{\today}
\begin{abstract}
Topological crystalline insulators (TCIs) are insulating materials whose topological property relies on generic crystalline symmetries.  Based on first-principles calculations, we study a three-dimensional (3D) crystal constructed by stacking two-dimensional TCI layers. Depending on the inter-layer interaction, the layered crystal can realize diverse 3D topological phases characterized by two mirror Chern numbers (MCNs) ($\mu_1,\mu_2$) defined on inequivalent mirror-invariant planes in the Brillouin zone. As an example, we demonstrate that new TCI phases can be realized in layered materials such as a PbSe (001) monolayer/h-BN heterostructure and can be tuned by mechanical strain.  Our results shed light on the role of the MCNs on inequivalent mirror-symmetric planes in reciprocal space and open new possibilities for finding new topological materials.
\end{abstract}
\maketitle
\acrodef{MCN}{mirror Chern number}
\acrodef{TCI}{topological crystalline insulator}
\acrodef{TI}{topological insulator}
\acrodef{2D}{two-dimensional}
\acrodef{3D}{three-dimensional}
\acrodef{BZ}{Brillouin zone}
\acrodef{DFT}{density functional theory}
New topological states of matter, \acp{TCI} \cite{Fu11p106802}, have been identified that extend the topological classification beyond the prototypical $Z_2$ classification based on time reversal symmetry \cite{Hasan10p3045, Qi11p1057}.  In \acp{TCI}, topological properties of electronic structure such as the presence of robust metallic surface states arise from crystal symmetries instead of time-reversal symmetries.  There are many proposed \ac{TCI} phases depending on different crystal symmetries \cite{Fang12p115112, Slager13p98, Chiu13p075142, Fang13p035119, Alexandradinata14p116403, Shiozaki14p165114}, yet those relying on mirror symmetry \cite{Hsieh12p982} are of particular interest as they have been experimentally observed in, for example, IV-VI semiconductors SnTe, ${\textrm{Pb}}_{1-x}{\textrm{Sn}}_x{\textrm{Te}}$, and ${\textrm{Pb}}_{1-x}{\textrm{Sn}}_x{\textrm{Se}}$ \cite{Dziawa12p1023, Liang13p2696, Tanaka13p235126, Okada13p5163, Xu12p1192}. More materials are theoretically proposed to realize the \ac{TCI} phases such as rocksalt semiconductors \cite{Sun13p235122, Tang14p041409}, pyrochlore iridates \cite{Kargarian13p156403}, graphene systems \cite{Kindermann13p1}, heavy fermion compounds \cite{Weng14p016403,Ye13p1}, and antiperovskites \cite{Hsieh14p081112}, including \ac{2D} materials such as SnTe thin films \cite{Liu14p178, Ozawa14p045309} and a (001) monolayer of PbSe \cite{Wrasse14p5717}.

Mirror-symmetric \ac{TCI}s are mathematically characterized by \acp{MCN}. The \ac{MCN} is a topological invariant defined by $\mu_1 \equiv (\mu_+-\mu_-)/2$ where $\mu_+$ and $\mu_-$ are Chern numbers of Bloch states with the opposite eigenvalues of a mirror operator ($M_z$) calculated on the mirror-invariant plane at $k_z=0$ in the \ac{BZ}. In a \ac{3D} crystal, there is a second \ac{MCN} ($\mu_2$) defined on the mirror-invariant plane at the boundary of the \ac{BZ} $k_z = \pi$ (in units of $1/a$, where $a$ is the length of the primitive lattice vector along the $z$-axis) \cite{Teo08p045426}. Moreover, considering different mirror symmetries, multiple pairs of \acp{MCN} ($\mu_1,\mu_2$) can be simultaneously present in three dimensions. A complete characterization of \ac{3D} \acp{TCI} requires consideration of all the \acp{MCN}, which may allow for the possibility of new states of matter, where \acp{MCN} are locked together or undergo separate transitions.  Nonetheless, previous study based only on $\mu_1$ has not explored this situation.

In this paper, by considering \acp{MCN} on all inequivalent mirror-symmetric planes in reciprocal space, we study new topological states of matter realized in a 3D layered crystal generated by stacking \ac{2D} \ac{TCI} layers. We show that the layered system realizes a new class of \ac{3D} \ac{TCI}s when inter-layer interaction is weak, which we will refer to as a layered \ac{TCI}. The layered TCI is characterized by equal and nonzero first and second \acp{MCN} $\mu_1 = \mu_2 \ne 0$ with a number of metallic surface states eqaul to $|\mu_1| + |\mu_2|$.  Increasing the inter-layer interaction, we then show that the layered \ac{TCI} undergoes topological phase transitions that change the \acp{MCN} $(\mu_1,\mu_2)$.  Based on first-principles calculations, we predict that a heterostructure consisting of alternating layers of PbSe monolayer and hexagonal BN (h-BN) sheet realizes the layered \ac{TCI} indexed by (2,2), and that it undergoes distinct topological phase transitions in the sequence ($\mu_1,\mu_2$): $(2,2) \rightarrow (0,2) \rightarrow (0,0)$ under external uniaxial tensile strain. Our findings shed light on new states of matter allowed by the presence of multiple \acp{MCN} in a \ac{3D} crystal. They may also help guide the discovery of more topological materials.

\begin{figure}[t] 
\includegraphics[width=0.50\textwidth]{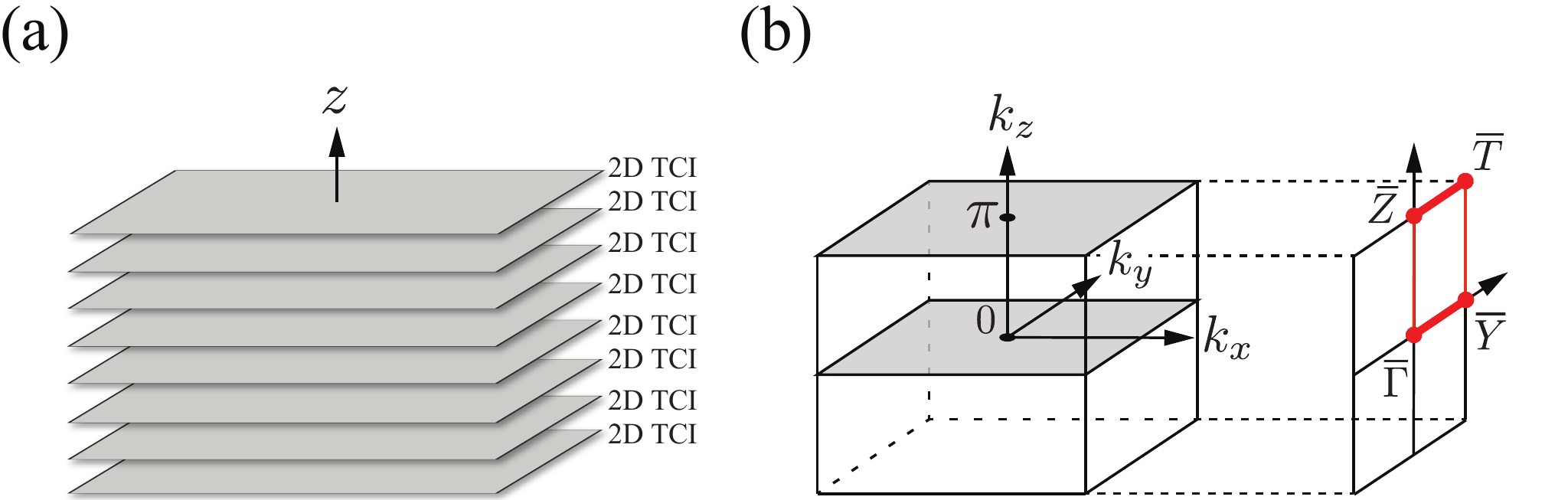}
\caption{\label{fig1} Schematic drawing of  (a) the proposed layered structure consisting of a stack of \ac{2D} \ac{TCI} layers and (b) the corresponding \ac{BZ}. Only eight periods of the crystal are shown in the crystal structure. The \ac{2D} planes at $k_z = 0$ and $k_z = \pi$ in the \ac{BZ} (gray shaded) are mirror-invariant under the reflection , $z \rightarrow -z$, on which the first and second mirror Chern numbers (MCNs) are defined, respectively.  For a surface normal (100), those mirror-invariant planes are projected on $\overline{Y} - \overline{\Gamma} - \overline{Y}$ and $\overline{T} -  \overline{Z} - \overline{T}$, respectively, along which surface Dirac points are expected to occur.  }
\end{figure}
Before presenting the results, we first briefly explain how \ac{2D} \ac{TCI} layers with a non-zero \ac{MCN} $\mu_{\mathrm 2D} = n$ ($n\ne0$) can be stacked into a new class of \ac{3D} \ac{TCI}s characterized by $\mu_1 = \mu_2 = n$. Consider first a layered system consisting of \ac{2D} \acp{TCI} with $\mu_{\mathrm 2D} = n \ne 0$ stacked along the normal direction to the plane (defined as $z$-direction) as shown in FIG.\ \ref{fig1}(a). The layered system then respects the mirror symmetry $M_z$ that defines the \ac{MCN} of the \ac{2D} \ac{TCI} $\mu_{\mathrm 2D}$ in the plane of each layer. Now, let us initially assume that the interaction between the layers is negligibly weak, so that every cross section of the \ac{3D} \ac{BZ} at constant $k_z$ is essentially a copy of the \ac{2D} \ac{BZ} of the film. In particular, the mirror invariant planes at $k_z = 0$ and $k_z = \pi$ [See Fig.\ \ref{fig1}(b)] should adopt the same \ac{MCN} as the \ac{2D} \ac{TCI}, and thus be indexed by $(\mu_1,\mu_2) = (n,n)$. For mirror symmetries inequivalent to $M_z$ (if any), the corresponding \acp{MCN} are all trivial (0,0) because the  mirror planes allowed by the layered geometry are normal to the films, and the crystalline surfaces respecting the mirror symmetries are essentially the \ac{2D} \acp{TCI} without metallic (surface) states. This means that the proposed \ac{TCI} are characterized by the coupled \acp{MCN} ($n,n$) for $M_z$ and ($0,0$) for any mirror symmetry inequivalent to $M_z$. Turning on the inter-layer interaction in a way that respects the mirror symmetry, the \acp{MCN} should persist within a finite range of the interaction, until the system experiences a topological phase transition through a gap closure \cite{Smith11p056401}, which can lead either to a new topological state where the indices are decoupled or to a conventional insulating state. 

We demonstrate the topological phases associated with \acp{MCN} $(\mu_1,\mu_2)$ and their transitions from first principles by applying the above theory to a PbSe/h-BN heterostructure.  Our calculation is performed with \ac{DFT} including the Perdew-Burke-Ernzerhof \cite{Perdew96p3865} generalized gradient approximation as implemented in the QUANTUM ESPRESSO package \cite{Giannozzi09p395502}.  The atomic potentials are modeled by norm-conserving, optimized, designed nonlocal pseudopotentials with fully relativistic spin-orbit interaction generated by the OPIUM package \cite{Rappe90p1227, Ramer99p12471}. The wave functions are expanded in a plane-wave basis with an energy cutoff of 650 eV. For computational convenience, the energy cutoff is reduced to 540 eV when calculating the surface band structure of PbSe(001) monolayers.  The van der Waals interaction is described based on the semiempirical dispersion-correction DFT (DFT-D) method \cite{Grimme06p1787}. The tight-binding model, introduced in Ref. \cite{Hsieh12p982}, is also employed to analyze the \ac{DFT} results on (001) PbSe layers using parameter sets obtained from our \ac{DFT} calculations. The results are consistent with the previous studies of IV-VI semiconductors, including PbSe \cite{Hsieh12p982, Liu14p178, Wrasse14p5717}.  The unit cell of the PbSe/h-BN heterostructure is generated by contracting the in-plane lattice constants of the h-BN sheet so that they match the pristine lattice constant of the PbSe.  We have checked that the artificial contraction has negligible influence on the electronic structure near the Fermi energy, as h-BN has a wide band gap.

\begin{figure}[t] 
\includegraphics[width=0.47\textwidth]{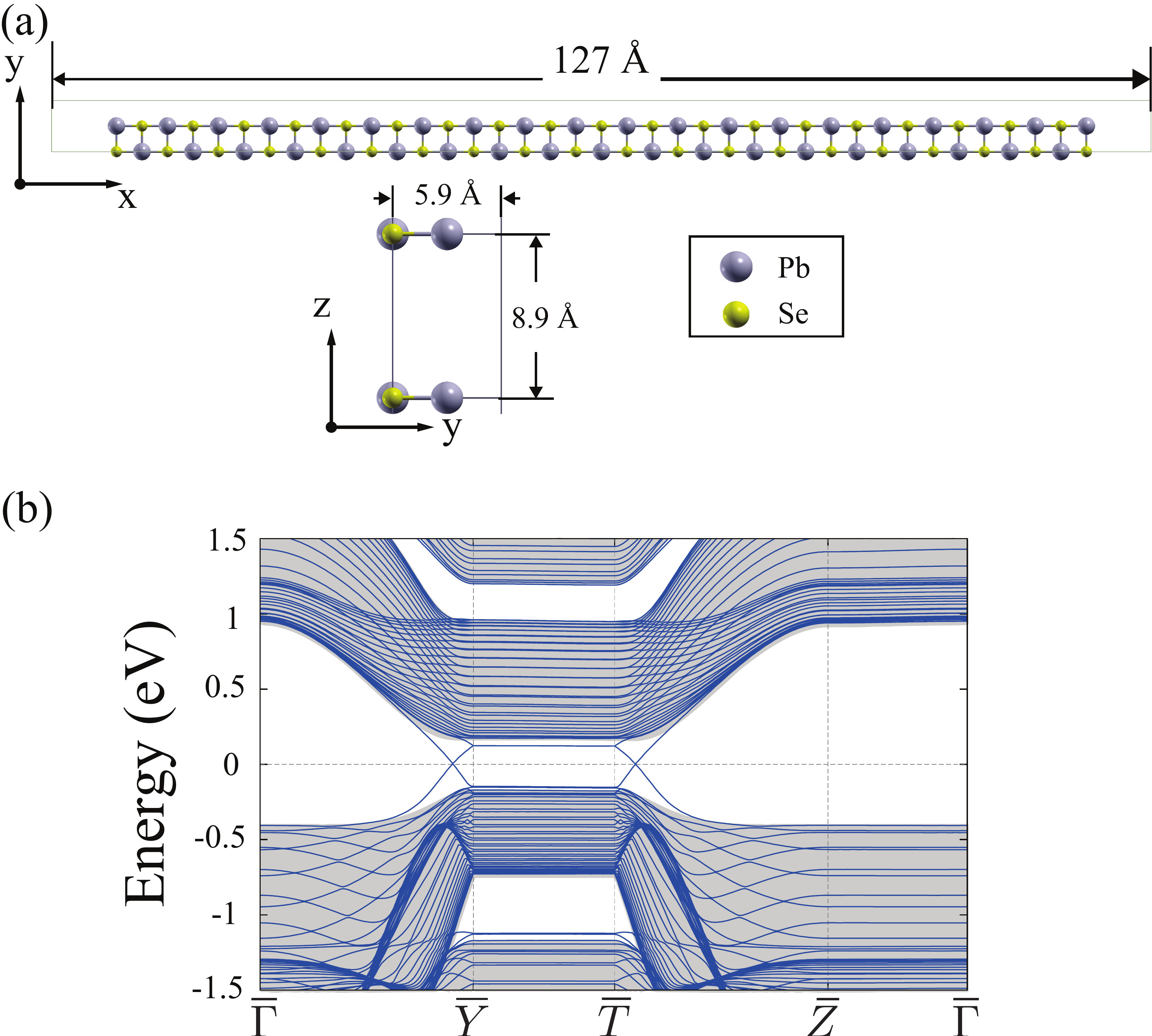}
\caption{\label{fig2} (Color online) Atomic geometry and band structure for a slab of PbSe multilayers with a (100) face. (a) The top view (upper) and lateral view (left lower) of the slab supercell.  (b) The band structure for the slab geometry indexed in the \ac{2D} \ac{BZ}. The \acp{MCN} on the $k_z = 0$ and $k_z = \pi$ planes each give rise to the \ac{2D} Dirac cone on one of the mirror-symmetric lines. The gray shaded regions represent the surface-projected bulk continuum bands.  }
\end{figure}
We first build a layered \ac{TCI} based on (001) PbSe monolayers. Whereas PbSe is a trivial insulator in a \ac{3D} rocksalt geometry, (001) PbSe monolayer is expected to be a \ac{2D} \ac{TCI}, indexed by the \ac{MCN} $|\mu_{\mathrm 2D}| = 2 $ \cite{Wrasse14p5717}.  As shown in Fig.\ \ref{fig2}(a), we consider a system consisting of PbSe monolayers stacked along the perpendicular direction to the plane ([001]-direction), so that Pb (and Se) atoms form chains along [001], separated by 8.9 \AA. The other crystal parameters are set to those of the bulk PbSe. In this way, the inter-layer interaction remains weak, and the resulting system is a layered \ac{TCI} indexed by $(2,2)$ associated with the (001) mirror plane. The system respects additional mirror symmetries about \{100\} and \{110\} mirror planes, on which the \acp{MCN} are all trivial as discussed above.

The calculated \acp{MCN} $(2,2)$ signal the presence of four surface states on the mirror-symmetric facets. As depicted in Fig.\ \ref{fig1}(b), for a surface containing $k_z$, the surface \ac{BZ} has two inequivalent mirror-symmetric lines, $\overline{Y}$ - $\overline{\Gamma}$ - $\overline{Y}$ and $\overline{Z}$ - $\overline{T}$ - $\overline{Z}$ which are the projections of the $0$ and $\pi$ mirror-planes into the surface plane.  The absolute values of the \acp{MCN} $|\mu_1|$ and $|\mu_2|$ dictate the numbers of pairs of counter-propagating surface states on the $k_z = 0$ and $k_z = \pi$ mirror lines, respectively.  It follows that there must exist two pairs of surface states along each line.  To look for the surface states guaranteed by the \acp{MCN}, we calculate the \ac{2D} band structure for the slab geometry illustrated in Fig.\ \ref{fig2}(a). Figure \ref{fig2}(b) shows the band structure for a slab exposing the (100) surface to vacuum along high symmetric lines [See Fig.\ \ref{fig1}(b)]. As expected, in addition to the bulk states in the gray region, we find surface states that traverse the gap forming Dirac points on the mirror-symmetric lines $\overline{\Gamma}$ - $\overline{Y}$ and $\overline{T}$ - $\overline{Z}$. It is clear from the results that the (100) surface has four total Dirac points [two shown in Fig.\ \ref{fig2}(b) and two more at the minus of these], dictated by their sum of the absolute value $|\mu_1|+|\mu_2|$. Note that $\overline{Y}$ - $\overline{T}$ and $\overline{Z}$ - $\overline{\Gamma}$ host no metallic surface states. This proves $(\mu_1,\mu_2) = (0,0)$ on (010) mirror planes. 

\begin{figure}[t] 
  \includegraphics[width=0.47\textwidth]{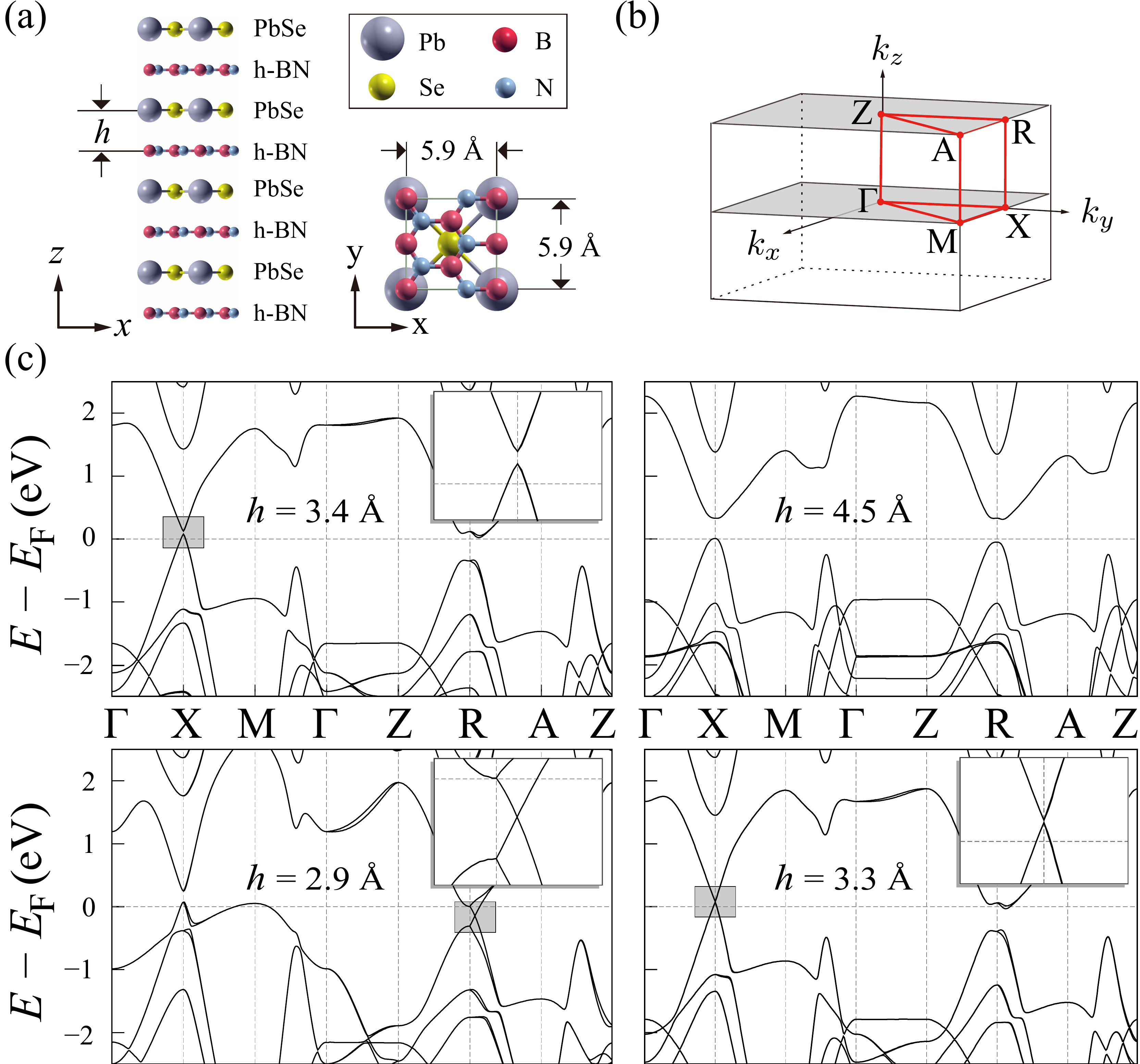}
  \caption{\label{fig3}
(Color online) Topological phase transition in the (001) PbSe/h-BN heterostructure. (a) The crystal structure and (b) the corresponding \ac{BZ}. (c) The band structures of the PbSe/h-BN heterostructure at $h = 3.4$ \AA, $h = 4.5$ \AA, $h = 2.9$ \AA, and $h = 3.3$ \AA. The \acp{MCN} hosted in $k_z = 0$ and $k_z = \pi$ mirror planes at the equilibrium inter-layer distance $h = 3.4$ \AA\, are adiabatically the same as those at $h > 3.4$, which is a layered \ac{TCI} with the \acp{MCN} $(\mu_1,\mu_2) = (2,2)$.  The Dirac cones at $h = 2.9$ \AA, and $h = 3.3$ \AA\ (magnified in the inset) respectively signal the topological phase transitions $(0,0) \rightarrow (0,2)$ and $(0,2) \rightarrow (2,2)$.  }
\end{figure}
Having demonstrated the layered \ac{TCI} with $\mu_1 = \mu_2 = 2$ in the layered PbSe system, we now consider a more realistic material. Above, we manually fixed the distance between the PbSe monolayers to 8.9 \AA\, to make the inter-layer interaction weak, but this is energetically unfavorable as the PbSe layers feel a repulsive force as the same cations and anions in different layers face each other at the interface. To stabilize the system, while keeping the inter-layer interaction weak, we insert a h-BN sheet which serves as spacer between the neighboring PbSe layers, as shown in Fig.\ \ref{fig3}(a). A h-BN sheet is a normal insulator with a wide band gap of 5 eV, which suggests that the band topology of the heterostructure should be governed by bands from the PbSe films. We find that the heterostructure has an equilibrium distance ($h_0$) of 3.4 \AA, with a binding energy of 0.08 eV per unit cell of PbSe, which indicates that the interaction is in the typical van der Waals regime. 

In Fig.\ \ref{fig3}, we show the band structures of the PbSe/h-BN heterostructure along the high-symmetric lines in the \ac{BZ} calculated for various PbSe-h-BN inter-layer distances ($h$).  First, at equilibrium $h_0$, the system is found to be semi-metallic with a small hole pocket at $X$ on the $k_z = 0$ plane and an electron pocket near $R$ on the $k_z = \pi$ plane. Then, by increasing the inter-layer distance from $h_0$ ($h > h_0$), the band gap at $X$ keeps increasing, and the system eventually becomes an insulator when $h > 3.5$ \AA. Increasing $h$ further, we find that the system remains insulating without closing the band gap. Thus, we assign the layered \ac{TCI} phase indexed by (2,2) to the system when $h > 3.5$.

\begin{figure}[t] 
  \includegraphics[width=0.48\textwidth]{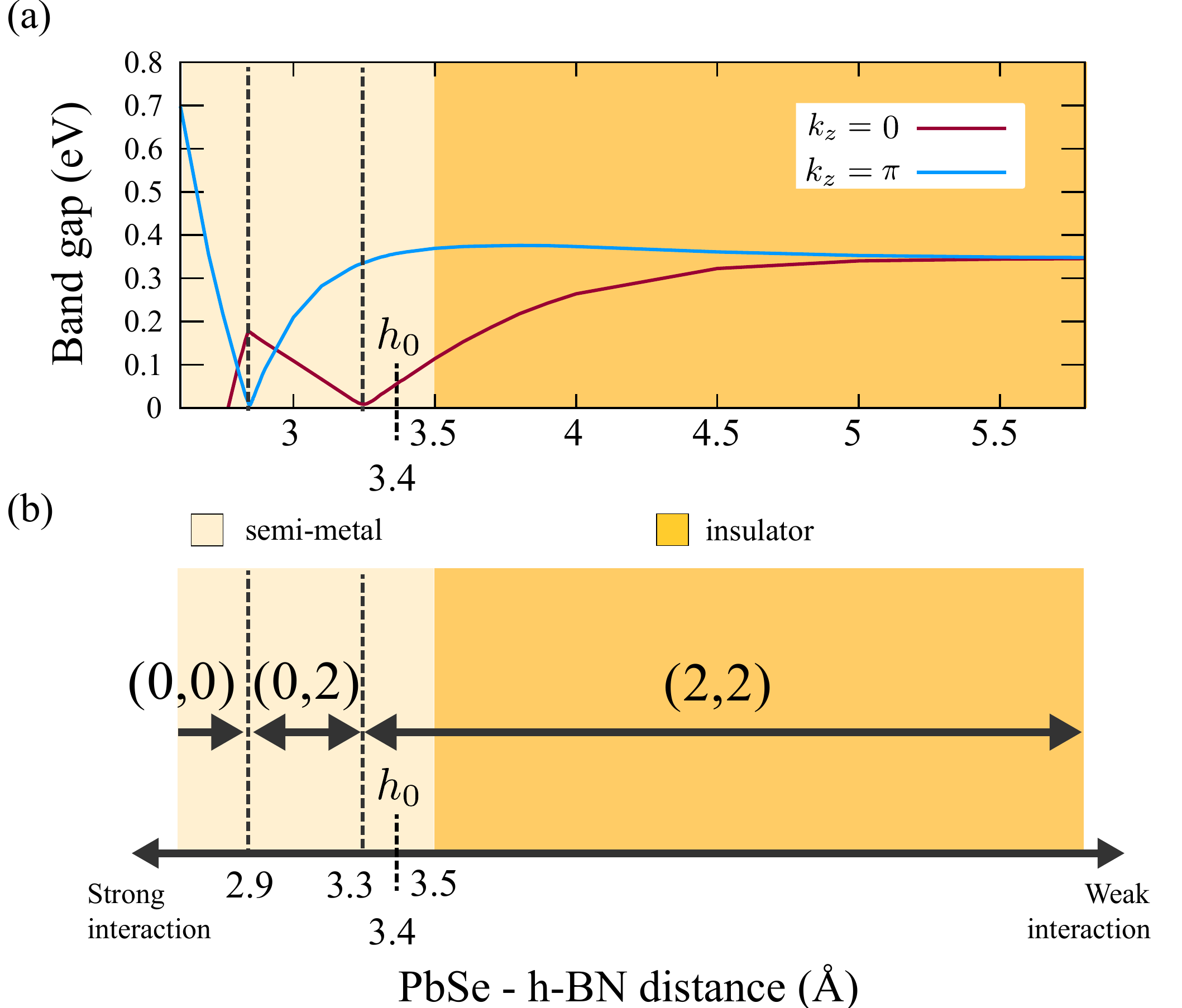}
  \caption{\label{fig4} (Color online) (a) Band gap evolution (lines for $k_z = 0$ and $k_z = \pi$ gap; shading for overall gap) as a function of the distance between PbSe and h-BN layers $h$ and (b) corresponding topological phase diagram for the (001) PbSe/h-BN heterostructure.  $h_0$ is the equilibrium position.  }
\end{figure}

Conversely, by decreasing the inter-layer distance from equilibrium $h < h_0$, which enhances the inter-layer interaction, we find that the system undergoes topological phase transitions signaled by the appearance of the Dirac points. As presented in Fig.\ \ref{fig3}(c), the Dirac points appear at $h = 3.25$ \AA \, and $h = 2.95 $ \AA \, on the $k_z = 0$ and $k_z = \pi$ mirror planes, respectively. The \acp{MCN}, calculated using all the valence bands on each mirror-symmetric plane, change from (2,2) to (0,2) and from (0,2) to (0,0) as $h$ passes through 3.24 \AA\,and $2.85$ \AA, respectively. All the topological phase transitions occur in a a region where the system is a semimetal because of overlapping bands.  Below $h = 2.77$ \AA, the valence band maximum becomes higher in energy than the conduction band minimum on the $k_z = 0 $ plane, so the \ac{MCN} is not defined.  Therefore, from the strong to weak inter-layer interaction regimes, four distinct topological phases appear, as shown in the phase diagram in Fig.\ \ref{fig4}(b); a trivial semimetal phase with (0,0), a topological semimetal phase with $(\mu_1,\mu_2) = (0,2)$, a topological semimetal phase with $(\mu_1,\mu_2) = (2,2)$, and the layered \ac{TCI} phase with $(\mu_1,\mu_2) = (2,2)$.  Although the heterostructure of PbSe/h-BN sheets is expected to be semi-metallic at ambient pressure, we expect that these phases should be accessible under mechanical strain including the proposed (2,2) layered \ac{TCI} or by inserting another h-BN sheet between PbSe layers.  We also expect that the phase transitions demonstrated in this system should be representative of layered \acp{TCI}, and heterostructures of \ac{2D} \acp{TCI} can be considered as hosts of diverse topological phases accessible by engineering the inter-layer interaction.  The calculated inter-layer distances may vary depending on details of the crystal geometry like a stacking registry between PbSe-h-BN layers, yet the qualitative features should remain intact. 

Finally, we note that layered \acp{TCI} are analogous to weak \acp{TI} \cite{Fu07p106803, Mong12p076804}.  Weak \acp{TI}, characterized by zero $Z_2$ invariant yet non-zero weak topological indices $(\nu_1 \nu_2 \nu_3)$, is essentially a stack of \ac{2D} \ac{TI} layers along the perpendicular direction that corresponds to ${\bf G} = \nu_1 {\bf b}_1 + \nu_2 {\bf b}_2 + \nu_3 {\bf b}_3$  in the BZ \cite{Fu07p106803}, having even numbers of robust Dirac cones at the surfaces perpendicular to the \ac{2D} \ac{TI} layers \cite{Ringel12p045102, Mong12p076804}.  Similarly, 3D \ac{TCI} with the same first and second mirror Chern numbers is like layered \ac{2D} \ac{TCI}s, having $|\mu_1| + |\mu_2|$ Dirac cones on the surfaces normal to the \ac{2D} \ac{TCI} layers.  Also, like weak topological indices, ($\mu_1,\mu_2$) are sensitive to the translational symmetry of the crystal. For instance, we find that a period-doubling along the $z$-axis changes the \acp{MCN} $(\mu_1,\mu_2)$: $(n,n) \rightarrow (2n,0)$, and $(n,-n) \rightarrow (0,2n)$ due to the \ac{BZ} folding, which can be induced by inter-layer bonding or a registry shift between 2D TCI layers.  Indeed, we have found the \acp{MCN} to be (4,0) when stacking the PbSe layers with a registry shift atoms between Pb and Se atoms in every layer.  Notwithstanding the sensitivity, it is important to note that the total number of surface Dirac cones on the mirror symmetric surfaces, dictated by $|\mu_1|+|\mu_2|$, is invariant under the period-doubling, and the second \ac{MCN} is thus indispensable for characterizing the layered \acp{TCI}, and indeed all 3D TCIs.

In conclusion, we propose new topological states of matter generated by stacking \ac{2D} \acp{TCI}, where simultaneous consideration of multiple \acp{MCN} is necessary.  In the non-interacting limit between layers, the layered TCI phase emerges where the first and second MCNs ($\mu_1,\mu_2$) are coupled.  The layered TCI is a generic class of 3D TCIs which can apply to a range of \ac{2D} \ac{TCI} materials. For example, a SnTe thin film, which is expected to be a \ac{2D} \ac{TCI} when cleaved into an odd number of (001) layers $\ge 5$ \cite{Liu14p178}, can play the role of the PbSe layer in the PbSe/h-BN heterostructure, thus realizing the layered TCI indexed by (2,2) when the layers are well separated. Apart from the non-interacting regime, we find topological semimetal phases indexed by $(2,2)$ and $(0,2)$ and trivial semimetal phase with $(0,0)$.  Despite the presence of metallic bulk states, the phase transitions should be observable via experimental techniques such as angle-resolved photoemission spectroscopy. Our findings shed light on the possibility of new \ac{TCI} phases relying on the fact that a crystal in three dimensions can have multiple MCNs hosted on inequivalent mirror planes in reciprocal lattice. These may open the way towards the search for new topological materials, based on which quantum devices for electronics as well as spintronics can be built.

\begin{acknowledgments}
We thank Fan Zhang and Steve M. Young for helpful discussion.  CLK acknowledges support as a Simons Investigator grant from the Simons Foundation.  EJM acknowledges support from the Department of Energy under Grant No. FG02-ER45118.  AMR acknowledges support from the Department of Energy Office of Basic Energy Sciences, under grant number DE-FG02-07ER15920.  Computational support is provided by the HPCMO of the U.S. DOD and the NERSC of the U.S. DOE.
\end{acknowledgments}

\bibliography{ref}

%merlin.mbs apsrev4-1.bst 2010-07-25 4.21a (PWD, AO, DPC) hacked
%Control: key (0)
%Control: author (8) initials jnrlst
%Control: editor formatted (1) identically to author
%Control: production of article title (-1) disabled
%Control: page (0) single
%Control: year (1) truncated
%Control: production of eprint (0) enabled
\begin{thebibliography}{35}%
\makeatletter
\providecommand \@ifxundefined [1]{%
 \@ifx{#1\undefined}
}%
\providecommand \@ifnum [1]{%
 \ifnum #1\expandafter \@firstoftwo
 \else \expandafter \@secondoftwo
 \fi
}%
\providecommand \@ifx [1]{%
 \ifx #1\expandafter \@firstoftwo
 \else \expandafter \@secondoftwo
 \fi
}%
\providecommand \natexlab [1]{#1}%
\providecommand \enquote  [1]{``#1''}%
\providecommand \bibnamefont  [1]{#1}%
\providecommand \bibfnamefont [1]{#1}%
\providecommand \citenamefont [1]{#1}%
\providecommand \href@noop [0]{\@secondoftwo}%
\providecommand \href [0]{\begingroup \@sanitize@url \@href}%
\providecommand \@href[1]{\@@startlink{#1}\@@href}%
\providecommand \@@href[1]{\endgroup#1\@@endlink}%
\providecommand \@sanitize@url [0]{\catcode `\\12\catcode `\$12\catcode
  `\&12\catcode `\#12\catcode `\^12\catcode `\_12\catcode `\%12\relax}%
\providecommand \@@startlink[1]{}%
\providecommand \@@endlink[0]{}%
\providecommand \url  [0]{\begingroup\@sanitize@url \@url }%
\providecommand \@url [1]{\endgroup\@href {#1}{\urlprefix }}%
\providecommand \urlprefix  [0]{URL }%
\providecommand \Eprint [0]{\href }%
\providecommand \doibase [0]{http://dx.doi.org/}%
\providecommand \selectlanguage [0]{\@gobble}%
\providecommand \bibinfo  [0]{\@secondoftwo}%
\providecommand \bibfield  [0]{\@secondoftwo}%
\providecommand \translation [1]{[#1]}%
\providecommand \BibitemOpen [0]{}%
\providecommand \bibitemStop [0]{}%
\providecommand \bibitemNoStop [0]{.\EOS\space}%
\providecommand \EOS [0]{\spacefactor3000\relax}%
\providecommand \BibitemShut  [1]{\csname bibitem#1\endcsname}%
\let\auto@bib@innerbib\@empty
%</preamble>
\bibitem [{\citenamefont {Fu}(2011)}]{Fu11p106802}%
  \BibitemOpen
  \bibfield  {author} {\bibinfo {author} {\bibfnamefont {L.}~\bibnamefont
  {Fu}},\ }\href {\doibase 10.1103/PhysRevLett.106.106802} {\bibfield
  {journal} {\bibinfo  {journal} {Phys. Rev. Lett.}\ }\textbf {\bibinfo
  {volume} {106}},\ \bibinfo {pages} {106802} (\bibinfo {year}
  {2011})}\BibitemShut {NoStop}%
\bibitem [{\citenamefont {Hasan}\ and\ \citenamefont
  {Kane}(2010)}]{Hasan10p3045}%
  \BibitemOpen
  \bibfield  {author} {\bibinfo {author} {\bibfnamefont {M.~Z.}\ \bibnamefont
  {Hasan}}\ and\ \bibinfo {author} {\bibfnamefont {C.~L.}\ \bibnamefont
  {Kane}},\ }\href {\doibase 10.1103/RevModPhys.82.3045} {\bibfield  {journal}
  {\bibinfo  {journal} {Rev. Mod. Phys.}\ }\textbf {\bibinfo {volume} {82}},\
  \bibinfo {pages} {3045} (\bibinfo {year} {2010})}\BibitemShut {NoStop}%
\bibitem [{\citenamefont {Qi}\ and\ \citenamefont {Zhang}(2011)}]{Qi11p1057}%
  \BibitemOpen
  \bibfield  {author} {\bibinfo {author} {\bibfnamefont {X.-L.}\ \bibnamefont
  {Qi}}\ and\ \bibinfo {author} {\bibfnamefont {S.-C.}\ \bibnamefont {Zhang}},\
  }\href {\doibase 10.1103/RevModPhys.83.1057} {\bibfield  {journal} {\bibinfo
  {journal} {Rev. Mod. Phys.}\ }\textbf {\bibinfo {volume} {83}},\ \bibinfo
  {pages} {1057} (\bibinfo {year} {2011})}\BibitemShut {NoStop}%
\bibitem [{\citenamefont {Fang}\ \emph {et~al.}(2012)\citenamefont {Fang},
  \citenamefont {Gilbert},\ and\ \citenamefont {Bernevig}}]{Fang12p115112}%
  \BibitemOpen
  \bibfield  {author} {\bibinfo {author} {\bibfnamefont {C.}~\bibnamefont
  {Fang}}, \bibinfo {author} {\bibfnamefont {M.~J.}\ \bibnamefont {Gilbert}}, \
  and\ \bibinfo {author} {\bibfnamefont {B.~A.}\ \bibnamefont {Bernevig}},\
  }\href {\doibase 10.1103/PhysRevB.86.115112} {\bibfield  {journal} {\bibinfo
  {journal} {Phys. Rev. B}\ }\textbf {\bibinfo {volume} {86}},\ \bibinfo
  {pages} {115112} (\bibinfo {year} {2012})}\BibitemShut {NoStop}%
\bibitem [{\citenamefont {Slager}\ \emph {et~al.}(2013)\citenamefont {Slager},
  \citenamefont {Mesaros}, \citenamefont {Juricic},\ and\ \citenamefont
  {Zaanen}}]{Slager13p98}%
  \BibitemOpen
  \bibfield  {author} {\bibinfo {author} {\bibfnamefont {R.-J.}\ \bibnamefont
  {Slager}}, \bibinfo {author} {\bibfnamefont {A.}~\bibnamefont {Mesaros}},
  \bibinfo {author} {\bibfnamefont {V.}~\bibnamefont {Juricic}}, \ and\
  \bibinfo {author} {\bibfnamefont {J.}~\bibnamefont {Zaanen}},\ }\href
  {\doibase 10.1038/nphys2513} {\bibfield  {journal} {\bibinfo  {journal}
  {Nature Physics}\ }\textbf {\bibinfo {volume} {9}},\ \bibinfo {pages} {98}
  (\bibinfo {year} {2013})}\BibitemShut {NoStop}%
\bibitem [{\citenamefont {Chiu}\ \emph {et~al.}(2013)\citenamefont {Chiu},
  \citenamefont {Yao},\ and\ \citenamefont {Ryu}}]{Chiu13p075142}%
  \BibitemOpen
  \bibfield  {author} {\bibinfo {author} {\bibfnamefont {C.-K.}\ \bibnamefont
  {Chiu}}, \bibinfo {author} {\bibfnamefont {H.}~\bibnamefont {Yao}}, \ and\
  \bibinfo {author} {\bibfnamefont {S.}~\bibnamefont {Ryu}},\ }\href {\doibase
  10.1103/PhysRevB.88.075142} {\bibfield  {journal} {\bibinfo  {journal} {Phys.
  Rev. B}\ }\textbf {\bibinfo {volume} {88}},\ \bibinfo {pages} {075142}
  (\bibinfo {year} {2013})}\BibitemShut {NoStop}%
\bibitem [{\citenamefont {Fang}\ \emph {et~al.}(2013)\citenamefont {Fang},
  \citenamefont {Gilbert},\ and\ \citenamefont {Bernevig}}]{Fang13p035119}%
  \BibitemOpen
  \bibfield  {author} {\bibinfo {author} {\bibfnamefont {C.}~\bibnamefont
  {Fang}}, \bibinfo {author} {\bibfnamefont {M.~J.}\ \bibnamefont {Gilbert}}, \
  and\ \bibinfo {author} {\bibfnamefont {B.~A.}\ \bibnamefont {Bernevig}},\
  }\href {\doibase 10.1103/PhysRevB.87.035119} {\bibfield  {journal} {\bibinfo
  {journal} {Phys. Rev. B}\ }\textbf {\bibinfo {volume} {87}},\ \bibinfo
  {pages} {035119} (\bibinfo {year} {2013})}\BibitemShut {NoStop}%
\bibitem [{\citenamefont {Alexandradinata}\ \emph {et~al.}(2014)\citenamefont
  {Alexandradinata}, \citenamefont {Fang}, \citenamefont {Gilbert},\ and\
  \citenamefont {Bernevig}}]{Alexandradinata14p116403}%
  \BibitemOpen
  \bibfield  {author} {\bibinfo {author} {\bibfnamefont {A.}~\bibnamefont
  {Alexandradinata}}, \bibinfo {author} {\bibfnamefont {C.}~\bibnamefont
  {Fang}}, \bibinfo {author} {\bibfnamefont {M.~J.}\ \bibnamefont {Gilbert}}, \
  and\ \bibinfo {author} {\bibfnamefont {B.~A.}\ \bibnamefont {Bernevig}},\
  }\href {\doibase 10.1103/PhysRevLett.113.116403} {\bibfield  {journal}
  {\bibinfo  {journal} {Phys. Rev. Lett.}\ }\textbf {\bibinfo {volume} {113}},\
  \bibinfo {pages} {116403} (\bibinfo {year} {2014})}\BibitemShut {NoStop}%
\bibitem [{\citenamefont {Shiozaki}\ and\ \citenamefont
  {Sato}(2014)}]{Shiozaki14p165114}%
  \BibitemOpen
  \bibfield  {author} {\bibinfo {author} {\bibfnamefont {K.}~\bibnamefont
  {Shiozaki}}\ and\ \bibinfo {author} {\bibfnamefont {M.}~\bibnamefont
  {Sato}},\ }\href {\doibase 10.1103/PhysRevB.90.165114} {\bibfield  {journal}
  {\bibinfo  {journal} {Phys. Rev. B}\ }\textbf {\bibinfo {volume} {90}},\
  \bibinfo {pages} {165114} (\bibinfo {year} {2014})}\BibitemShut {NoStop}%
\bibitem [{\citenamefont {Hsieh}\ \emph {et~al.}(2012)\citenamefont {Hsieh},
  \citenamefont {Lin}, \citenamefont {Liu}, \citenamefont {Duan}, \citenamefont
  {Bansil},\ and\ \citenamefont {Fu}}]{Hsieh12p982}%
  \BibitemOpen
  \bibfield  {author} {\bibinfo {author} {\bibfnamefont {T.~H.}\ \bibnamefont
  {Hsieh}}, \bibinfo {author} {\bibfnamefont {H.}~\bibnamefont {Lin}}, \bibinfo
  {author} {\bibfnamefont {J.}~\bibnamefont {Liu}}, \bibinfo {author}
  {\bibfnamefont {W.}~\bibnamefont {Duan}}, \bibinfo {author} {\bibfnamefont
  {A.}~\bibnamefont {Bansil}}, \ and\ \bibinfo {author} {\bibfnamefont
  {L.}~\bibnamefont {Fu}},\ }\href@noop {} {\bibfield  {journal} {\bibinfo
  {journal} {Nat. Commun.}\ }\textbf {\bibinfo {volume} {3}},\ \bibinfo {pages}
  {982} (\bibinfo {year} {2012})}\BibitemShut {NoStop}%
\bibitem [{\citenamefont {Dziawa}\ \emph {et~al.}(2012)\citenamefont {Dziawa},
  \citenamefont {Kowalski}, \citenamefont {Dybko}, \citenamefont {Buczko},
  \citenamefont {Szczerbakow}, \citenamefont {Szot}, \citenamefont
  {{\L}usakowska}, \citenamefont {Balasubramanian}, \citenamefont {Wojek},
  \citenamefont {Berntsen}, \citenamefont {Tjernberg},\ and\ \citenamefont
  {Story}}]{Dziawa12p1023}%
  \BibitemOpen
  \bibfield  {author} {\bibinfo {author} {\bibfnamefont {P.}~\bibnamefont
  {Dziawa}}, \bibinfo {author} {\bibfnamefont {B.~J.}\ \bibnamefont
  {Kowalski}}, \bibinfo {author} {\bibfnamefont {K.}~\bibnamefont {Dybko}},
  \bibinfo {author} {\bibfnamefont {R.}~\bibnamefont {Buczko}}, \bibinfo
  {author} {\bibfnamefont {A.}~\bibnamefont {Szczerbakow}}, \bibinfo {author}
  {\bibfnamefont {M.}~\bibnamefont {Szot}}, \bibinfo {author} {\bibfnamefont
  {E.}~\bibnamefont {{\L}usakowska}}, \bibinfo {author} {\bibfnamefont
  {T.}~\bibnamefont {Balasubramanian}}, \bibinfo {author} {\bibfnamefont
  {B.~M.}\ \bibnamefont {Wojek}}, \bibinfo {author} {\bibfnamefont {M.~H.}\
  \bibnamefont {Berntsen}}, \bibinfo {author} {\bibfnamefont {O.}~\bibnamefont
  {Tjernberg}}, \ and\ \bibinfo {author} {\bibfnamefont {T.}~\bibnamefont
  {Story}},\ }\href {\doibase 10.1038/nmat3449} {\bibfield  {journal} {\bibinfo
   {journal} {Nature Materials}\ }\textbf {\bibinfo {volume} {11}},\ \bibinfo
  {pages} {1023} (\bibinfo {year} {2012})}\BibitemShut {NoStop}%
\bibitem [{\citenamefont {Liang}\ \emph {et~al.}(2013)\citenamefont {Liang},
  \citenamefont {Gibson}, \citenamefont {Xiong}, \citenamefont {Hirschberger},
  \citenamefont {Koduvayur}, \citenamefont {Cava},\ and\ \citenamefont
  {Ong}}]{Liang13p2696}%
  \BibitemOpen
  \bibfield  {author} {\bibinfo {author} {\bibfnamefont {T.}~\bibnamefont
  {Liang}}, \bibinfo {author} {\bibfnamefont {Q.}~\bibnamefont {Gibson}},
  \bibinfo {author} {\bibfnamefont {J.}~\bibnamefont {Xiong}}, \bibinfo
  {author} {\bibfnamefont {M.}~\bibnamefont {Hirschberger}}, \bibinfo {author}
  {\bibfnamefont {S.~P.}\ \bibnamefont {Koduvayur}}, \bibinfo {author}
  {\bibfnamefont {R.~J.}\ \bibnamefont {Cava}}, \ and\ \bibinfo {author}
  {\bibfnamefont {N.~P.}\ \bibnamefont {Ong}},\ }\href@noop {} {\bibfield
  {journal} {\bibinfo  {journal} {Nat. Commun.}\ }\textbf {\bibinfo {volume}
  {4}},\ \bibinfo {pages} {2696} (\bibinfo {year} {2013})}\BibitemShut
  {NoStop}%
\bibitem [{\citenamefont {Tanaka}\ \emph {et~al.}(2013)\citenamefont {Tanaka},
  \citenamefont {Shoman}, \citenamefont {Nakayama}, \citenamefont {Souma},
  \citenamefont {Sato}, \citenamefont {Takahashi}, \citenamefont {Novak},
  \citenamefont {Segawa},\ and\ \citenamefont {Ando}}]{Tanaka13p235126}%
  \BibitemOpen
  \bibfield  {author} {\bibinfo {author} {\bibfnamefont {Y.}~\bibnamefont
  {Tanaka}}, \bibinfo {author} {\bibfnamefont {T.}~\bibnamefont {Shoman}},
  \bibinfo {author} {\bibfnamefont {K.}~\bibnamefont {Nakayama}}, \bibinfo
  {author} {\bibfnamefont {S.}~\bibnamefont {Souma}}, \bibinfo {author}
  {\bibfnamefont {T.}~\bibnamefont {Sato}}, \bibinfo {author} {\bibfnamefont
  {T.}~\bibnamefont {Takahashi}}, \bibinfo {author} {\bibfnamefont
  {M.}~\bibnamefont {Novak}}, \bibinfo {author} {\bibfnamefont
  {K.}~\bibnamefont {Segawa}}, \ and\ \bibinfo {author} {\bibfnamefont
  {Y.}~\bibnamefont {Ando}},\ }\href {\doibase 10.1103/PhysRevB.88.235126}
  {\bibfield  {journal} {\bibinfo  {journal} {Phys. Rev. B}\ }\textbf {\bibinfo
  {volume} {88}},\ \bibinfo {pages} {235126} (\bibinfo {year}
  {2013})}\BibitemShut {NoStop}%
\bibitem [{\citenamefont {Okada}\ \emph {et~al.}(2013)\citenamefont {Okada},
  \citenamefont {Serbyn}, \citenamefont {Lin}, \citenamefont {Walkup},
  \citenamefont {Zhou}, \citenamefont {Dhital}, \citenamefont {Neupane},
  \citenamefont {Xu}, \citenamefont {Wang}, \citenamefont {Sankar},
  \citenamefont {Chou}, \citenamefont {Bansil}, \citenamefont {Hasan},
  \citenamefont {Wilson}, \citenamefont {Fu},\ and\ \citenamefont
  {Madhavan}}]{Okada13p5163}%
  \BibitemOpen
  \bibfield  {author} {\bibinfo {author} {\bibfnamefont {Y.}~\bibnamefont
  {Okada}}, \bibinfo {author} {\bibfnamefont {M.}~\bibnamefont {Serbyn}},
  \bibinfo {author} {\bibfnamefont {H.}~\bibnamefont {Lin}}, \bibinfo {author}
  {\bibfnamefont {D.}~\bibnamefont {Walkup}}, \bibinfo {author} {\bibfnamefont
  {W.}~\bibnamefont {Zhou}}, \bibinfo {author} {\bibfnamefont {C.}~\bibnamefont
  {Dhital}}, \bibinfo {author} {\bibfnamefont {M.}~\bibnamefont {Neupane}},
  \bibinfo {author} {\bibfnamefont {S.}~\bibnamefont {Xu}}, \bibinfo {author}
  {\bibfnamefont {Y.~J.}\ \bibnamefont {Wang}}, \bibinfo {author}
  {\bibfnamefont {R.}~\bibnamefont {Sankar}}, \bibinfo {author} {\bibfnamefont
  {F.}~\bibnamefont {Chou}}, \bibinfo {author} {\bibfnamefont {A.}~\bibnamefont
  {Bansil}}, \bibinfo {author} {\bibfnamefont {M.~Z.}\ \bibnamefont {Hasan}},
  \bibinfo {author} {\bibfnamefont {S.~D.}\ \bibnamefont {Wilson}}, \bibinfo
  {author} {\bibfnamefont {L.}~\bibnamefont {Fu}}, \ and\ \bibinfo {author}
  {\bibfnamefont {V.}~\bibnamefont {Madhavan}},\ }\href {\doibase
  10.1126/science.1239451} {\bibfield  {journal} {\bibinfo  {journal}
  {Science}\ }\textbf {\bibinfo {volume} {341}},\ \bibinfo {pages} {1496}
  (\bibinfo {year} {2013})}\BibitemShut {NoStop}%
\bibitem [{\citenamefont {Xu}\ \emph {et~al.}(2012)\citenamefont {Xu},
  \citenamefont {Liu}, \citenamefont {Alidoust}, \citenamefont {Neupane},
  \citenamefont {Qian}, \citenamefont {Belopolski}, \citenamefont {Denlinger},
  \citenamefont {Wang}, \citenamefont {Lin}, \citenamefont {Wray},
  \citenamefont {Landolt}, \citenamefont {Slomski}, \citenamefont {Dil},
  \citenamefont {Marcinkova}, \citenamefont {Morosan}, \citenamefont {Gibson},
  \citenamefont {Sankar}, \citenamefont {Chou}, \citenamefont {Cava},
  \citenamefont {Bansil},\ and\ \citenamefont {Hasan}}]{Xu12p1192}%
  \BibitemOpen
  \bibfield  {author} {\bibinfo {author} {\bibfnamefont {S.-Y.}\ \bibnamefont
  {Xu}}, \bibinfo {author} {\bibfnamefont {C.}~\bibnamefont {Liu}}, \bibinfo
  {author} {\bibfnamefont {N.}~\bibnamefont {Alidoust}}, \bibinfo {author}
  {\bibfnamefont {M.}~\bibnamefont {Neupane}}, \bibinfo {author} {\bibfnamefont
  {D.}~\bibnamefont {Qian}}, \bibinfo {author} {\bibfnamefont {I.}~\bibnamefont
  {Belopolski}}, \bibinfo {author} {\bibfnamefont {J.~D.}\ \bibnamefont
  {Denlinger}}, \bibinfo {author} {\bibfnamefont {Y.~J.}\ \bibnamefont {Wang}},
  \bibinfo {author} {\bibfnamefont {H.}~\bibnamefont {Lin}}, \bibinfo {author}
  {\bibfnamefont {L.~A.}\ \bibnamefont {Wray}}, \bibinfo {author}
  {\bibfnamefont {G.}~\bibnamefont {Landolt}}, \bibinfo {author} {\bibfnamefont
  {B.}~\bibnamefont {Slomski}}, \bibinfo {author} {\bibfnamefont {J.~H.}\
  \bibnamefont {Dil}}, \bibinfo {author} {\bibfnamefont {A.}~\bibnamefont
  {Marcinkova}}, \bibinfo {author} {\bibfnamefont {E.}~\bibnamefont {Morosan}},
  \bibinfo {author} {\bibfnamefont {Q.}~\bibnamefont {Gibson}}, \bibinfo
  {author} {\bibfnamefont {R.}~\bibnamefont {Sankar}}, \bibinfo {author}
  {\bibfnamefont {F.~C.}\ \bibnamefont {Chou}}, \bibinfo {author}
  {\bibfnamefont {R.~J.}\ \bibnamefont {Cava}}, \bibinfo {author}
  {\bibfnamefont {A.}~\bibnamefont {Bansil}}, \ and\ \bibinfo {author}
  {\bibfnamefont {M.~Z.}\ \bibnamefont {Hasan}},\ }\href {\doibase
  10.1038/ncomms2191} {\bibfield  {journal} {\bibinfo  {journal} {Nat.
  Commun.}\ }\textbf {\bibinfo {volume} {3}},\ \bibinfo {pages} {1192}
  (\bibinfo {year} {2012})}\BibitemShut {NoStop}%
\bibitem [{\citenamefont {Sun}\ \emph {et~al.}(2013)\citenamefont {Sun},
  \citenamefont {Zhong}, \citenamefont {Shirakawa}, \citenamefont {Franchini},
  \citenamefont {Li}, \citenamefont {Li}, \citenamefont {Yunoki},\ and\
  \citenamefont {Chen}}]{Sun13p235122}%
  \BibitemOpen
  \bibfield  {author} {\bibinfo {author} {\bibfnamefont {Y.}~\bibnamefont
  {Sun}}, \bibinfo {author} {\bibfnamefont {Z.}~\bibnamefont {Zhong}}, \bibinfo
  {author} {\bibfnamefont {T.}~\bibnamefont {Shirakawa}}, \bibinfo {author}
  {\bibfnamefont {C.}~\bibnamefont {Franchini}}, \bibinfo {author}
  {\bibfnamefont {D.}~\bibnamefont {Li}}, \bibinfo {author} {\bibfnamefont
  {Y.}~\bibnamefont {Li}}, \bibinfo {author} {\bibfnamefont {S.}~\bibnamefont
  {Yunoki}}, \ and\ \bibinfo {author} {\bibfnamefont {X.-Q.}\ \bibnamefont
  {Chen}},\ }\href {\doibase 10.1103/PhysRevB.88.235122} {\bibfield  {journal}
  {\bibinfo  {journal} {Phys. Rev. B}\ }\textbf {\bibinfo {volume} {88}},\
  \bibinfo {pages} {235122} (\bibinfo {year} {2013})}\BibitemShut {NoStop}%
\bibitem [{\citenamefont {Tang}\ \emph {et~al.}(2014)\citenamefont {Tang},
  \citenamefont {Yan}, \citenamefont {Cao}, \citenamefont {Wu}, \citenamefont
  {Felser},\ and\ \citenamefont {Duan}}]{Tang14p041409}%
  \BibitemOpen
  \bibfield  {author} {\bibinfo {author} {\bibfnamefont {P.}~\bibnamefont
  {Tang}}, \bibinfo {author} {\bibfnamefont {B.}~\bibnamefont {Yan}}, \bibinfo
  {author} {\bibfnamefont {W.}~\bibnamefont {Cao}}, \bibinfo {author}
  {\bibfnamefont {S.-C.}\ \bibnamefont {Wu}}, \bibinfo {author} {\bibfnamefont
  {C.}~\bibnamefont {Felser}}, \ and\ \bibinfo {author} {\bibfnamefont
  {W.}~\bibnamefont {Duan}},\ }\href {\doibase 10.1103/PhysRevB.89.041409}
  {\bibfield  {journal} {\bibinfo  {journal} {Phys. Rev. B}\ }\textbf {\bibinfo
  {volume} {89}},\ \bibinfo {pages} {041409} (\bibinfo {year}
  {2014})}\BibitemShut {NoStop}%
\bibitem [{\citenamefont {Kargarian}\ and\ \citenamefont
  {Fiete}(2013)}]{Kargarian13p156403}%
  \BibitemOpen
  \bibfield  {author} {\bibinfo {author} {\bibfnamefont {M.}~\bibnamefont
  {Kargarian}}\ and\ \bibinfo {author} {\bibfnamefont {G.~A.}\ \bibnamefont
  {Fiete}},\ }\href@noop {} {\bibfield  {journal} {\bibinfo  {journal} {Phys.
  Rev. Lett.}\ }\textbf {\bibinfo {volume} {110}},\ \bibinfo {pages} {156403}
  (\bibinfo {year} {2013})}\BibitemShut {NoStop}%
\bibitem [{\citenamefont {Kindermann}(2013)}]{Kindermann13p1}%
  \BibitemOpen
  \bibfield  {author} {\bibinfo {author} {\bibfnamefont {M.}~\bibnamefont
  {Kindermann}},\ }\href@noop {} {\  (\bibinfo {year} {2013})},\ \Eprint
  {http://arxiv.org/abs/arXiv:1309.1667} {arXiv:1309.1667} \BibitemShut
  {NoStop}%
\bibitem [{\citenamefont {Weng}\ \emph {et~al.}(2014)\citenamefont {Weng},
  \citenamefont {Zhao}, \citenamefont {Wang}, \citenamefont {Fang},\ and\
  \citenamefont {Dai}}]{Weng14p016403}%
  \BibitemOpen
  \bibfield  {author} {\bibinfo {author} {\bibfnamefont {H.}~\bibnamefont
  {Weng}}, \bibinfo {author} {\bibfnamefont {J.}~\bibnamefont {Zhao}}, \bibinfo
  {author} {\bibfnamefont {Z.}~\bibnamefont {Wang}}, \bibinfo {author}
  {\bibfnamefont {Z.}~\bibnamefont {Fang}}, \ and\ \bibinfo {author}
  {\bibfnamefont {X.}~\bibnamefont {Dai}},\ }\href {\doibase
  10.1103/PhysRevLett.112.016403} {\bibfield  {journal} {\bibinfo  {journal}
  {Phys. Rev. Lett.}\ }\textbf {\bibinfo {volume} {112}},\ \bibinfo {pages}
  {016403} (\bibinfo {year} {2014})}\BibitemShut {NoStop}%
\bibitem [{\citenamefont {Ye}\ \emph {et~al.}(2013)\citenamefont {Ye},
  \citenamefont {Allen},\ and\ \citenamefont {Sun}}]{Ye13p1}%
  \BibitemOpen
  \bibfield  {author} {\bibinfo {author} {\bibfnamefont {M.}~\bibnamefont
  {Ye}}, \bibinfo {author} {\bibfnamefont {J.~W.}\ \bibnamefont {Allen}}, \
  and\ \bibinfo {author} {\bibfnamefont {K.}~\bibnamefont {Sun}},\ }\href@noop
  {} {\  (\bibinfo {year} {2013})},\ \Eprint
  {http://arxiv.org/abs/arXiv:1307.7191} {arXiv:1307.7191} \BibitemShut
  {NoStop}%
\bibitem [{\citenamefont {Hsieh}\ \emph {et~al.}(2014)\citenamefont {Hsieh},
  \citenamefont {Liu},\ and\ \citenamefont {Fu}}]{Hsieh14p081112}%
  \BibitemOpen
  \bibfield  {author} {\bibinfo {author} {\bibfnamefont {T.~H.}\ \bibnamefont
  {Hsieh}}, \bibinfo {author} {\bibfnamefont {J.}~\bibnamefont {Liu}}, \ and\
  \bibinfo {author} {\bibfnamefont {L.}~\bibnamefont {Fu}},\ }\href {\doibase
  10.1103/PhysRevB.90.081112} {\bibfield  {journal} {\bibinfo  {journal} {Phys.
  Rev. B}\ }\textbf {\bibinfo {volume} {90}},\ \bibinfo {pages} {081112}
  (\bibinfo {year} {2014})}\BibitemShut {NoStop}%
\bibitem [{\citenamefont {Liu}\ \emph {et~al.}(2014)\citenamefont {Liu},
  \citenamefont {Hsieh}, \citenamefont {Wei}, \citenamefont {Duan},
  \citenamefont {Moodera},\ and\ \citenamefont {Fu}}]{Liu14p178}%
  \BibitemOpen
  \bibfield  {author} {\bibinfo {author} {\bibfnamefont {J.}~\bibnamefont
  {Liu}}, \bibinfo {author} {\bibfnamefont {T.~H.}\ \bibnamefont {Hsieh}},
  \bibinfo {author} {\bibfnamefont {P.}~\bibnamefont {Wei}}, \bibinfo {author}
  {\bibfnamefont {W.}~\bibnamefont {Duan}}, \bibinfo {author} {\bibfnamefont
  {J.}~\bibnamefont {Moodera}}, \ and\ \bibinfo {author} {\bibfnamefont
  {L.}~\bibnamefont {Fu}},\ }\href@noop {} {\bibfield  {journal} {\bibinfo
  {journal} {Nature Materials}\ }\textbf {\bibinfo {volume} {13}},\ \bibinfo
  {pages} {178} (\bibinfo {year} {2014})}\BibitemShut {NoStop}%
\bibitem [{\citenamefont {Ozawa}\ \emph {et~al.}(2014)\citenamefont {Ozawa},
  \citenamefont {Yamakage}, \citenamefont {Sato},\ and\ \citenamefont
  {Tanaka}}]{Ozawa14p045309}%
  \BibitemOpen
  \bibfield  {author} {\bibinfo {author} {\bibfnamefont {H.}~\bibnamefont
  {Ozawa}}, \bibinfo {author} {\bibfnamefont {A.}~\bibnamefont {Yamakage}},
  \bibinfo {author} {\bibfnamefont {M.}~\bibnamefont {Sato}}, \ and\ \bibinfo
  {author} {\bibfnamefont {Y.}~\bibnamefont {Tanaka}},\ }\href {\doibase
  10.1103/PhysRevB.90.045309} {\bibfield  {journal} {\bibinfo  {journal} {Phys.
  Rev. B}\ }\textbf {\bibinfo {volume} {90}},\ \bibinfo {pages} {045309}
  (\bibinfo {year} {2014})}\BibitemShut {NoStop}%
\bibitem [{\citenamefont {Wrasse}\ and\ \citenamefont
  {Schmidt}(2014)}]{Wrasse14p5717}%
  \BibitemOpen
  \bibfield  {author} {\bibinfo {author} {\bibfnamefont {E.~O.}\ \bibnamefont
  {Wrasse}}\ and\ \bibinfo {author} {\bibfnamefont {T.~M.}\ \bibnamefont
  {Schmidt}},\ }\href {\doibase 10.1021/nl502481f} {\bibfield  {journal}
  {\bibinfo  {journal} {Nano Lett.}\ }\textbf {\bibinfo {volume} {14}},\
  \bibinfo {pages} {5717} (\bibinfo {year} {2014})}\BibitemShut {NoStop}%
\bibitem [{\citenamefont {Teo}\ \emph {et~al.}(2008)\citenamefont {Teo},
  \citenamefont {Fu},\ and\ \citenamefont {Kane}}]{Teo08p045426}%
  \BibitemOpen
  \bibfield  {author} {\bibinfo {author} {\bibfnamefont {J.~C.~Y.}\
  \bibnamefont {Teo}}, \bibinfo {author} {\bibfnamefont {L.}~\bibnamefont
  {Fu}}, \ and\ \bibinfo {author} {\bibfnamefont {C.~L.}\ \bibnamefont
  {Kane}},\ }\href {\doibase 10.1103/PhysRevB.78.045426} {\bibfield  {journal}
  {\bibinfo  {journal} {Phys. Rev. B}\ }\textbf {\bibinfo {volume} {78}},\
  \bibinfo {pages} {045426} (\bibinfo {year} {2008})}\BibitemShut {NoStop}%
\bibitem [{\citenamefont {Smith}\ \emph {et~al.}(2011)\citenamefont {Smith},
  \citenamefont {Banerjee}, \citenamefont {Pardo},\ and\ \citenamefont
  {Pickett}}]{Smith11p056401}%
  \BibitemOpen
  \bibfield  {author} {\bibinfo {author} {\bibfnamefont {J.~C.}\ \bibnamefont
  {Smith}}, \bibinfo {author} {\bibfnamefont {S.}~\bibnamefont {Banerjee}},
  \bibinfo {author} {\bibfnamefont {V.}~\bibnamefont {Pardo}}, \ and\ \bibinfo
  {author} {\bibfnamefont {W.~E.}\ \bibnamefont {Pickett}},\ }\href {\doibase
  10.1103/PhysRevLett.106.056401} {\bibfield  {journal} {\bibinfo  {journal}
  {Phys. Rev. Lett.}\ }\textbf {\bibinfo {volume} {106}},\ \bibinfo {pages}
  {056401} (\bibinfo {year} {2011})}\BibitemShut {NoStop}%
\bibitem [{\citenamefont {Perdew}\ \emph {et~al.}(1996)\citenamefont {Perdew},
  \citenamefont {Burke},\ and\ \citenamefont {Ernzerhof}}]{Perdew96p3865}%
  \BibitemOpen
  \bibfield  {author} {\bibinfo {author} {\bibfnamefont {J.~P.}\ \bibnamefont
  {Perdew}}, \bibinfo {author} {\bibfnamefont {K.}~\bibnamefont {Burke}}, \
  and\ \bibinfo {author} {\bibfnamefont {M.}~\bibnamefont {Ernzerhof}},\ }\href
  {\doibase 10.1103/PhysRevLett.77.3865} {\bibfield  {journal} {\bibinfo
  {journal} {Phys. Rev. Lett.}\ }\textbf {\bibinfo {volume} {77}},\ \bibinfo
  {pages} {3865} (\bibinfo {year} {1996})}\BibitemShut {NoStop}%
\bibitem [{\citenamefont {Giannozzi}\ \emph {et~al.}(2009)\citenamefont
  {Giannozzi}, \citenamefont {Baroni}, \citenamefont {Bonini}, \citenamefont
  {Calandra}, \citenamefont {Car}, \citenamefont {Cavazzoni}, \citenamefont
  {Ceresoli}, \citenamefont {Chiarotti}, \citenamefont {Cococcioni},
  \citenamefont {Dabo}, \citenamefont {Corso}, \citenamefont {de~Gironcoli},
  \citenamefont {Fabris}, \citenamefont {Fratesi}, \citenamefont {Gebauer},
  \citenamefont {Gerstmann}, \citenamefont {Gougoussis}, \citenamefont
  {Kokalj}, \citenamefont {Lazzeri}, \citenamefont {Martin-Samos},
  \citenamefont {Marzari}, \citenamefont {Mauri}, \citenamefont {Mazzarello},
  \citenamefont {Paolini}, \citenamefont {Pasquarello}, \citenamefont
  {Paulatto}, \citenamefont {Sbraccia}, \citenamefont {Scandolo}, \citenamefont
  {Sclauzero}, \citenamefont {Seitsonen}, \citenamefont {Smogunov},
  \citenamefont {Umari},\ and\ \citenamefont
  {Wentzcovitch}}]{Giannozzi09p395502}%
  \BibitemOpen
  \bibfield  {author} {\bibinfo {author} {\bibfnamefont {P.}~\bibnamefont
  {Giannozzi}}, \bibinfo {author} {\bibfnamefont {S.}~\bibnamefont {Baroni}},
  \bibinfo {author} {\bibfnamefont {N.}~\bibnamefont {Bonini}}, \bibinfo
  {author} {\bibfnamefont {M.}~\bibnamefont {Calandra}}, \bibinfo {author}
  {\bibfnamefont {R.}~\bibnamefont {Car}}, \bibinfo {author} {\bibfnamefont
  {C.}~\bibnamefont {Cavazzoni}}, \bibinfo {author} {\bibfnamefont
  {D.}~\bibnamefont {Ceresoli}}, \bibinfo {author} {\bibfnamefont {G.~L.}\
  \bibnamefont {Chiarotti}}, \bibinfo {author} {\bibfnamefont {M.}~\bibnamefont
  {Cococcioni}}, \bibinfo {author} {\bibfnamefont {I.}~\bibnamefont {Dabo}},
  \bibinfo {author} {\bibfnamefont {A.~D.}\ \bibnamefont {Corso}}, \bibinfo
  {author} {\bibfnamefont {S.}~\bibnamefont {de~Gironcoli}}, \bibinfo {author}
  {\bibfnamefont {S.}~\bibnamefont {Fabris}}, \bibinfo {author} {\bibfnamefont
  {G.}~\bibnamefont {Fratesi}}, \bibinfo {author} {\bibfnamefont
  {R.}~\bibnamefont {Gebauer}}, \bibinfo {author} {\bibfnamefont
  {U.}~\bibnamefont {Gerstmann}}, \bibinfo {author} {\bibfnamefont
  {C.}~\bibnamefont {Gougoussis}}, \bibinfo {author} {\bibfnamefont
  {A.}~\bibnamefont {Kokalj}}, \bibinfo {author} {\bibfnamefont
  {M.}~\bibnamefont {Lazzeri}}, \bibinfo {author} {\bibfnamefont
  {L.}~\bibnamefont {Martin-Samos}}, \bibinfo {author} {\bibfnamefont
  {N.}~\bibnamefont {Marzari}}, \bibinfo {author} {\bibfnamefont
  {F.}~\bibnamefont {Mauri}}, \bibinfo {author} {\bibfnamefont
  {R.}~\bibnamefont {Mazzarello}}, \bibinfo {author} {\bibfnamefont
  {S.}~\bibnamefont {Paolini}}, \bibinfo {author} {\bibfnamefont
  {A.}~\bibnamefont {Pasquarello}}, \bibinfo {author} {\bibfnamefont
  {L.}~\bibnamefont {Paulatto}}, \bibinfo {author} {\bibfnamefont
  {C.}~\bibnamefont {Sbraccia}}, \bibinfo {author} {\bibfnamefont
  {S.}~\bibnamefont {Scandolo}}, \bibinfo {author} {\bibfnamefont
  {G.}~\bibnamefont {Sclauzero}}, \bibinfo {author} {\bibfnamefont {A.~P.}\
  \bibnamefont {Seitsonen}}, \bibinfo {author} {\bibfnamefont {A.}~\bibnamefont
  {Smogunov}}, \bibinfo {author} {\bibfnamefont {P.}~\bibnamefont {Umari}}, \
  and\ \bibinfo {author} {\bibfnamefont {R.~M.}\ \bibnamefont {Wentzcovitch}},\
  }\href@noop {} {\bibfield  {journal} {\bibinfo  {journal} {J. Phys.: Condens.
  Matter.}\ }\textbf {\bibinfo {volume} {21}},\ \bibinfo {pages} {395502}
  (\bibinfo {year} {2009})}\BibitemShut {NoStop}%
\bibitem [{\citenamefont {Rappe}\ \emph {et~al.}(1990)\citenamefont {Rappe},
  \citenamefont {Rabe}, \citenamefont {Kaxiras},\ and\ \citenamefont
  {Joannopoulos}}]{Rappe90p1227}%
  \BibitemOpen
  \bibfield  {author} {\bibinfo {author} {\bibfnamefont {A.~M.}\ \bibnamefont
  {Rappe}}, \bibinfo {author} {\bibfnamefont {K.~M.}\ \bibnamefont {Rabe}},
  \bibinfo {author} {\bibfnamefont {E.}~\bibnamefont {Kaxiras}}, \ and\
  \bibinfo {author} {\bibfnamefont {J.~D.}\ \bibnamefont {Joannopoulos}},\
  }\href {\doibase 10.1103/PhysRevB.41.1227} {\bibfield  {journal} {\bibinfo
  {journal} {Phys. Rev. B}\ }\textbf {\bibinfo {volume} {41}},\ \bibinfo
  {pages} {1227} (\bibinfo {year} {1990})}\BibitemShut {NoStop}%
\bibitem [{\citenamefont {Ramer}\ and\ \citenamefont
  {Rappe}(1999)}]{Ramer99p12471}%
  \BibitemOpen
  \bibfield  {author} {\bibinfo {author} {\bibfnamefont {N.~J.}\ \bibnamefont
  {Ramer}}\ and\ \bibinfo {author} {\bibfnamefont {A.~M.}\ \bibnamefont
  {Rappe}},\ }\href {\doibase 10.1103/PhysRevB.59.12471} {\bibfield  {journal}
  {\bibinfo  {journal} {Phys. Rev. B}\ }\textbf {\bibinfo {volume} {59}},\
  \bibinfo {pages} {12471} (\bibinfo {year} {1999})}\BibitemShut {NoStop}%
\bibitem [{\citenamefont {Grimme}(2006)}]{Grimme06p1787}%
  \BibitemOpen
  \bibfield  {author} {\bibinfo {author} {\bibfnamefont {S.}~\bibnamefont
  {Grimme}},\ }\href {\doibase 10.1002/jcc.20495} {\bibfield  {journal}
  {\bibinfo  {journal} {J. Comput. Chem.}\ }\textbf {\bibinfo {volume} {27}},\
  \bibinfo {pages} {1787} (\bibinfo {year} {2006})}\BibitemShut {NoStop}%
\bibitem [{\citenamefont {Fu}\ \emph {et~al.}(2007)\citenamefont {Fu},
  \citenamefont {Kane},\ and\ \citenamefont {Mele}}]{Fu07p106803}%
  \BibitemOpen
  \bibfield  {author} {\bibinfo {author} {\bibfnamefont {L.}~\bibnamefont
  {Fu}}, \bibinfo {author} {\bibfnamefont {C.~L.}\ \bibnamefont {Kane}}, \ and\
  \bibinfo {author} {\bibfnamefont {E.~J.}\ \bibnamefont {Mele}},\ }\href
  {\doibase 10.1103/PhysRevLett.98.106803} {\bibfield  {journal} {\bibinfo
  {journal} {Phys. Rev. Lett.}\ }\textbf {\bibinfo {volume} {98}},\ \bibinfo
  {pages} {106803} (\bibinfo {year} {2007})}\BibitemShut {NoStop}%
\bibitem [{\citenamefont {Mong}\ \emph {et~al.}(2012)\citenamefont {Mong},
  \citenamefont {Bardarson},\ and\ \citenamefont {Moore}}]{Mong12p076804}%
  \BibitemOpen
  \bibfield  {author} {\bibinfo {author} {\bibfnamefont {R.~S.~K.}\
  \bibnamefont {Mong}}, \bibinfo {author} {\bibfnamefont {J.~H.}\ \bibnamefont
  {Bardarson}}, \ and\ \bibinfo {author} {\bibfnamefont {J.~E.}\ \bibnamefont
  {Moore}},\ }\href {\doibase 10.1103/PhysRevLett.108.076804} {\bibfield
  {journal} {\bibinfo  {journal} {Phys. Rev. Lett.}\ }\textbf {\bibinfo
  {volume} {108}},\ \bibinfo {pages} {076804} (\bibinfo {year}
  {2012})}\BibitemShut {NoStop}%
\bibitem [{\citenamefont {Ringel}\ \emph {et~al.}(2012)\citenamefont {Ringel},
  \citenamefont {Kraus},\ and\ \citenamefont {Stern}}]{Ringel12p045102}%
  \BibitemOpen
  \bibfield  {author} {\bibinfo {author} {\bibfnamefont {Z.}~\bibnamefont
  {Ringel}}, \bibinfo {author} {\bibfnamefont {Y.~E.}\ \bibnamefont {Kraus}}, \
  and\ \bibinfo {author} {\bibfnamefont {A.}~\bibnamefont {Stern}},\ }\href
  {\doibase 10.1103/PhysRevB.86.045102} {\bibfield  {journal} {\bibinfo
  {journal} {Phys. Rev. B}\ }\textbf {\bibinfo {volume} {86}},\ \bibinfo
  {pages} {045102} (\bibinfo {year} {2012})}\BibitemShut {NoStop}%
\end{thebibliography}%

\end{document}